\documentclass[showpacs,10pt,twocolumn,prl]{revtex4-1}

\usepackage{amsmath}
\usepackage{amssymb}
\usepackage{graphicx}
\usepackage{amssymb}
\usepackage{graphics}
\usepackage{epsfig}
\usepackage{CJK}
\usepackage{color}
\usepackage{soul}

\begin{document}

\begin{CJK*}{GBK}{Song}
\title{Anomalous Hall effect in weak-itinerant ferrimagnet FeCr$_2$Te$_4$}
\author{Yu Liu,$^{1,*}$ Hengxin Tan,$^{2}$ Zhixiang Hu,$^{1,3}$ Binghai Yan,$^{2}$ and C. Petrovic$^{1,3}$}
\affiliation{$^{1}$Department of Condensed Matter Physics and Materials Science, Brookhaven National Laboratory, Upton, New York 11973, USA\\
$^{2}$Department of Condensed Matter Physics, Weizmann Institute of Science, Rehovot 7610001, Israel\\
$^{3}$Department of Materials Science and Chemical Engineering, Stony Brook University, Stony Brook, NY 11790, USA}
\date{\today}

\begin{abstract}
We carried out a comprehensive study of electronic transport, thermal and thermodynamic properties in FeCr$_2$Te$_4$ single crystals. It exhibits bad-metallic behavior and anomalous Hall effect (AHE) below a weak-itinerant paramagentic-to-ferrimagnetic transition $T_c$ $\sim$ 123 K. The linear scaling between the anomalous Hall resistivity $\rho_{xy}$ and the longitudinal resistivity $\rho_{xx}$ implies that the AHE in FeCr$_2$Te$_4$ is most likely dominated by extrinsic skew-scattering mechanism rather than intrinsic KL or extrinsic side-jump mechanism, which is supported by our Berry phase calculations.
\end{abstract}
\maketitle
\end{CJK*}

\section{INTRODUCTION}

The anomalous Hall effect (AHE) in metals is linked to an asymmetry in carrier paths and the effects of spin-orbit interaction. This is typically observed in ferromagnets since an electric current induces a transverse voltage drop in zero magnetic field which is proportional to magnetization \cite{Hall,OnoseY}. Spin-orbit coupling in the ferromagnetic bands leads to anomalous carrier velocities and intrinsic AHE \cite{Karplus}. The intrinsic Kaplus-Luttinger (KL) mechanism can be reinterpreted as a manifestation of Berry-phase effects on occupied electronic Bloch states \cite{JungwirthT,Onoda}. The extrinsic mechanisms involving skew-scattering and side-jump mechanisms can also give rise to the AHE and are induced by asymmetric scattering of conduction electrons \cite{Smit,Berger}. In recent years it has been shown that the AHE velocities arise from the topological Berry curvature which generate an effective magnetic field in momentum space in varieties of Dirac materials with noncollinear spin configuration \cite{NagaosaN,XiaoD,ChenH,AsabaT,SinovaJ}.

FeCr$_2$Ch$_4$ (Ch = O, S, Se, Te) materials show rich correlated electron physics. FeCr$_{2}$O$_{4}$ spinel shows a complex magnetic phase diagram with a ferrimagnetic (FIM) and multiferroic order below 80 K, a strong spin-lattice coupling and orbital order due to the Jahn-Teller distortion \cite{Shirane,Tomiyasu,Singh,Maignan,Tsuda}. FeCr$_2$S$_4$ is a multiferroic ferrimagnet below $T_c$ = 165 K with large changes of resistivity in magnetic field \cite{Bertinshaw,Lin,Tsurkan,Tsurkan1,Ramirez}. FeCr$_2$Se$_4$ orders antiferromagnetically with $T_N$ = 218 K in an insulating state despite with larger ligand chalcogen atom \cite{Min,Snyder,OK}. FeCr$_2$S$_4$ and FeCr$_2$Se$_4$ have similar electronic structure with nearly trivalent Cr$^{3+}$ and divalent Fe$^{2+}$ states, and there is a strong hybridization between Fe $3d$- and Ch p-states \cite{Kang}. FeCr$_{2}$Te$_{4}$ shows no semiconducting gap and a FIM order below $T_c$ = 123 K \cite{Yadav,LIUYu}.

In this work, we performed a comprehensive study of electronic and thermal transport properties in FeCr$_{2}$Te$_{4}$ single crystals. The AHE observed below $T_c$ is dominated by the skew-scattering mechanism, i.e., by the Bloch state transport lifetime arising from electron scattering by impurities or defects in the presence of spin-orbit effects, and is smaller than the intrinsic AHE revealed by density functional calculations.

\section{EXPERIMENTAL AND COMPUTATIONAL DETAILS}

Single crystals growth and crystal structure details are described in ref.\cite{LIUYu}. Electrical and thermal transport were measured in quantum design PPMS-9. The longitudinal and Hall resistivity were measured using a standard four-probe method. In order to effectively eliminate the longitudinal resistivity contribution due to voltage probe misalignment, the Hall resistivity was obtained by the difference of transverse resistance measured at positive and negative fields, i.e., $\rho_{xy}(\mu_0H) = [\rho(+\mu_0H)-\rho(-\mu_0H)]/2$. Isothermal magnetization was measured in quantum design MPMS-XL5.

We performed density functional theory (DFT) calculations with the Perdew-Burke-Ernzerhof (PBE) \cite{PBE} exchange-correlation functional that is implemented in the Vienna ab initio simulation package (VASP) \cite{vasp2}. We adopted the experimental crystal structure with the ferrimagnetism (parallel to the lattice vector $c$) \cite{LIUYu}. The cutoff energy for the plane wave basis is 300 eV. A $k$-mesh of $10\times10\times10$ was used in the Brillouin zone sampling. The spin-orbit coupling was included. The intrinsic anomalous Hall conductivity (AHC) and Seebeck coefficient was calculated in a tight-binding scheme based on the maximally localized Wannier functions \cite{MLWF}.

\section{RESULTS AND DISCUSSIONS}

Figure 1(a) shows the temperature-dependent heat capacity $C_p(T)$ for FeCr$_2$Te$_4$. A clear anomaly around 123 K corresponds well to the paramagnetic (PM)-FIM transition. The high temperature $C_p(T)$ approaches the Dulong Petit value of $3NR$ $\approx$ 172 J mol$^{-1}$ K$^{-1}$, where $R$ = 8.314 J mol$^{-1}$ K$^{-1}$ is the molar gas constant. The low temperature data from 2 to 18 K are featureless and could be fitted by using $C_p(T)/T = \gamma + \beta T^2$, where the first term is the Sommerfeld electronic specific heat coefficient and the second term is low-temperature limit of lattice heat capacity [inset in Fig. 1(a)]. The fitting gives $\gamma$ = 61(2) mJ mol$^{-1}$ K$^{-2}$ and $\beta$ = 1.7(1) mJ mol$^{-1}$ K$^{-4}$. The Debye temperature $\Theta_D$ = 199(1) K can be calculated by using $\Theta_D = (12\pi^4NR/5\beta)^{1/3}$, where $N=7$ is the number of atoms per formula unit.

\begin{figure}
\centerline{\includegraphics[scale=1]{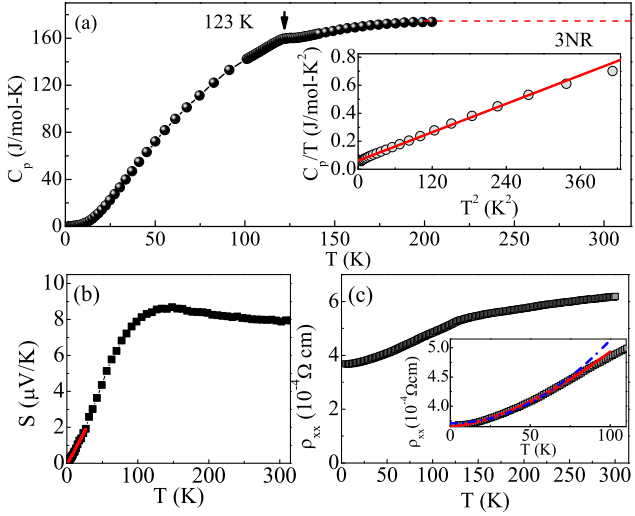}}
\caption{(Color online) (a) Temperature-dependent heat capacity $C_p(T)$ for FeCr$_2$Te$_4$. Inset shows the low temperature $C_p(T)/T$ vs $T^2$ curve fitted by $C_p(T)/T = \gamma + \beta T^2$. (b) Seebeck coefficient $S(T)$ and (c) in-plane resistivity $\rho_{xx}(T)$ for FeCr$_2$Te$_4$ single crystal. Inset in (c) shows data below 100 K fitted by $\rho(T) = \rho_0 + aT^{3/2} + bT^2$ (solid line) in comparison with $\rho(T) = \rho_0 + cT^2$ (dashed line).}
\label{MTH}
\end{figure}

The Seebeck coefficient $S(T)$ of FeCr$_2$Te$_4$ is positive in the whole temperature range, indicating dominant hole-type carriers [Fig. 1(b)]. The $S(T)$ changes slope around $T_c$ and gradually decreases with decreasing temperature. As we know, the $S(T)$ depends sensitively on the Fermi surface. The slope change of $S(T)$ reflects the possible reconstruction of Fermi surface passing through the PM-FIM transition. At low temperature, the diffusive Seebeck response of Fermi liquid dominates and is expected to be linear in $T$. In a metal with dominant single-band transport, the Seebeck coefficient could be described by the Mott relationship,
\begin{equation}
S = \frac{\pi^2}{3}\frac{k_B^2T}{e}\frac{N(\varepsilon_F)}{n},
\end{equation}
where $N(\varepsilon_F)$ is the density of states (DOS), $\varepsilon_F$ is the Fermi energy, $n$ is carrier concentration, $k_B$ is the Boltzman constant and $e$ is the absolute value of electronic charge \cite{Barnard}. The derived $dS/dT$ below 26 K is $\sim$ 0.074(1) $\mu$V K$^{-2}$. The $S(T)$ curve is consistent with our calculations based on Boltzmann equations and DFT band structure [see below in Fig. 4(b)]. The electronic specific heat is:
\begin{equation}
C_e=\frac{\pi^2}{3}k_B^2TN(\varepsilon_F).
\end{equation}
From Eq. (1), thermopower probes the specific heat per electron: $S = C_e/ne$. The units are V K$^{-1}$ for $S$, J K$^{-1}$ m$^{-3}$ for $C_e$, and m$^{-3}$ for $n$, respectively. It is common to express $\gamma = C_e/T$ in J K$^{-2}$ mol$^{-1}$ units. In order to focus on the $S/C_e$ ratio, we define a dimensionless quantity
\begin{equation}
q=\frac{S}{T}\frac{N_Ae}{\gamma},
\end{equation}
where $N_A$ is the Avogadro number. This gives the number of carriers per formula unit (proportional to $1/n$) \cite{Behnia}. The obtained $q$ = 0.10(1) indicates about 0.1 hole per formula unit within the Boltzmann framework \cite{Behnia}.

\begin{figure}
\centerline{\includegraphics[scale=0.4]{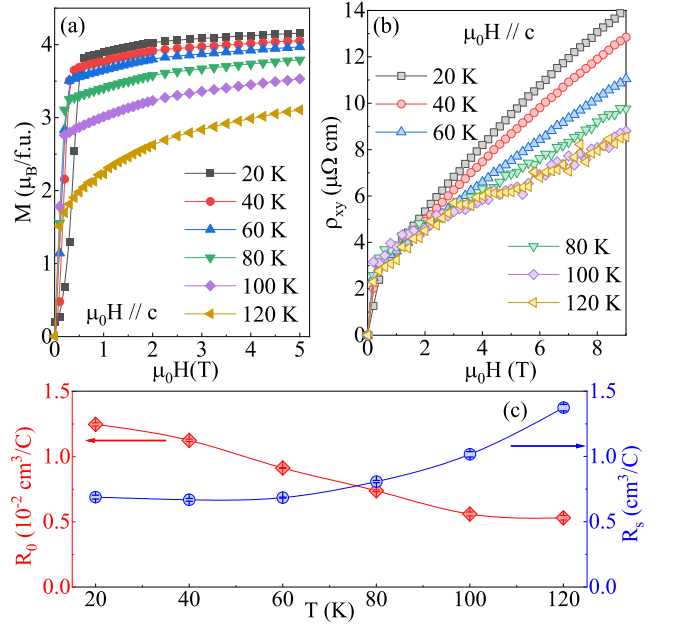}}
\caption{(Color online) Out-of-plane field dependence of (a) dc magnetization $M(\mu_0H)$ and (b) Hall resistivity $\rho_{xy}(\mu_0H)$ for FeCr$_2$Te$_4$ at indicated temperatures. (c) Temperature dependence of ordinary Hall coefficient $R_0$ (left axis) and anomalous Hall coefficient $R_s$ (right axis) fitted from the $\rho_{xy}$ vs $\mu_0H$ curves using $\rho_{xy} = R_0\mu_0H + R_sM$.}
\label{Arrot}
\end{figure}

Figure 1(c) shows the temperature-dependent in-plane resistivity $\rho_{xx}(T)$ of FeCr$_2$Te$_4$, indicating a metallic behavior with a relatively low residual resistivity ratio [RRR = $\rho$(300 K)/$\rho$(2 K) = 1.7]. A clear kink is observed at $T_c$, corresponding well to the PM-FIM transition. The renormalized spin fluctuation theory suggests that the electrical resistivity shows a $T^2$ dependence for itinerant ferromagnetic system \cite{Ueda1}. In FeCr$_2$Te$_4$, the low temperature resistivity fitting gives a better result by adding an additional $T^{3/2}$ term that describes the contribution of spin fluctuation scattering \cite{Rosch}.
\begin{equation}
\rho(T) = \rho_0 + aT^\frac{3}{2} + bT^2,
\end{equation}
where $\rho_0$ is the residual resistivity, $a$ and $b$ are constants. The fitting yields $\rho_0$ = 366(1) $\mu\Omega$ cm, $a$ = 1.00(3)$\times$$10^{-1}$ $\mu\Omega$ cm K$^{-1}$, and $b$ = 2.8(3)$\times$10$^{-3}$ $\mu\Omega$ cm K$^{-2}$, indicating the $T^{3/2}$ term predominates. This means the interaction between conduction electrons and localized spins could not be simply treated as a small perturbation to a system of free electrons, i.e., strong electron correlation should be considered in FeCr$_2$Te$_4$.

\begin{figure}
\centerline{\includegraphics[scale=1]{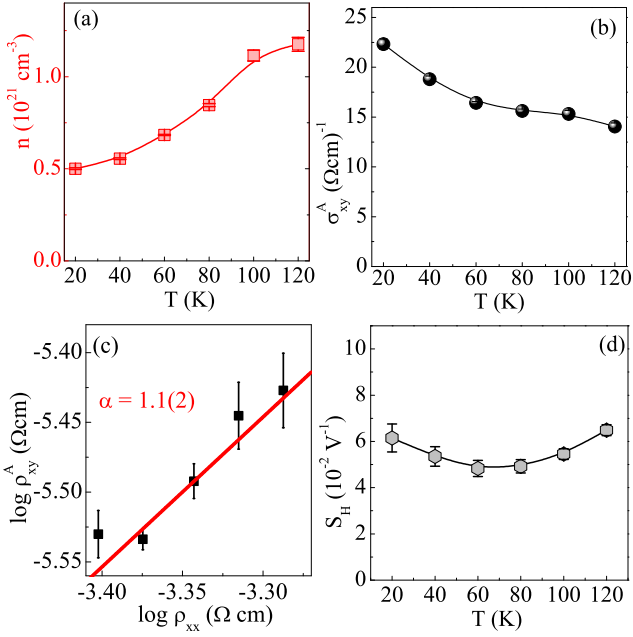}}
\caption{(Color online) Temperature dependence of the carrier concentration (a) and the anomalous Hall conductivity $\sigma_{xy}^A = \rho_{xy}^A / (\rho_{xx}^2+\rho_{xy}^2)$ (b). Scaling behavior of the anomalous Hall resistivity (c) and the coefficient $S_H = \mu_0R_s/\rho_{xx}^2$ (d).}
\label{KF}
\end{figure}

Figure 2(a) shows the isothermal magnetization measured at various temperatures below $T_c$. All the $M(\mu_0H)$ curves rapidly increase in low field and change slowly in high field. Field dependence of Hall resistivity $\rho_{xy}(\mu_0H)$ for FeCr$_2$Te$_4$ at the corresponding temperatures are depicted in Fig. 2(b). All the $\rho_{xy}(\mu_0H)$ curves jump in low field and then become linear-in-field in high field, indicating an AHE in FeCr$_2$Te$_4$ crystal. In general, the Hall resistivity $\rho_{xy}$ in ferromagnets is made up of two parts,
\begin{equation}
\rho_{xy} = \rho_{xy}^O + \rho_{xy}^A = R_0\mu_0H + R_sM,
\end{equation}
where $\rho_{xy}^O$ and $\rho_{xy}^A$ are the ordinary and anomalous Hall resistivity, respectively \cite{Wang,Yan, WangY,Onoda2008}. $R_0$ is the ordinary Hall coefficient from which apparent carrier concentration and type can be determined ($R_0 = 1/nq$). $R_s$ is the anomalous Hall coefficient. With a linear fit of $\rho_{xy}(\mu_0H)$ in high field, the slope and intercept corresponds to $R_0$ and $\rho_{xy}^A$, respectively. $R_s$ can be obtained from $\rho_{xy}^A = R_sM_s$ with $M_s$ taken from linear fit of $M(\mu_0H)$ curves in high field. The temperature dependence of derived $R_0$ and $R_s$ is plotted in Fig. 2(c). The value of $R_0$ is positive, in line with the positive $S(T)$, confirming the hole-type carries. The derived $R_s$ gradually decreases with decreasing temperature. Its magnitude is about two orders larger than that of $R_0$.

\begin{figure}
\centerline{\includegraphics[scale=1]{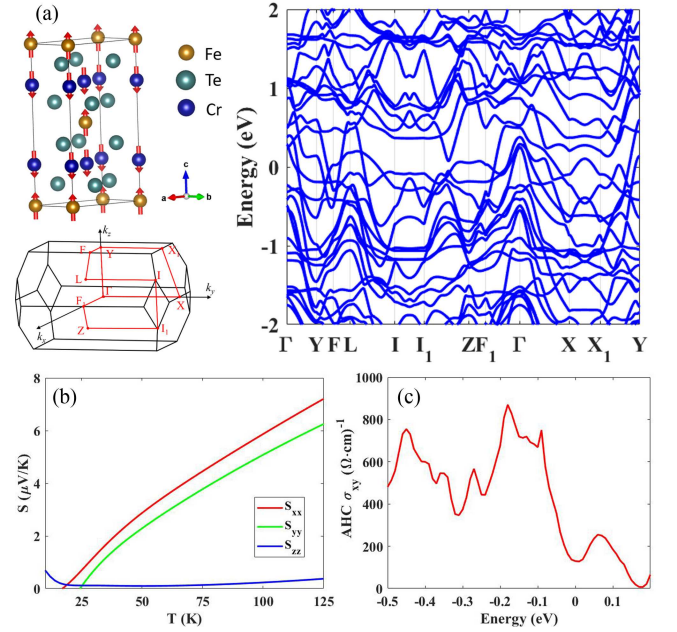}}
\caption{(Color online) (a) Crystal structure, Brillouin zone (BZ), and electronic structure of FeCr$_2$Te$_4$. The red vectors in the crystal structure represent the directions of the magnetic moments on Fe and Cr. The high symmetric k-paths in the BZ are shown. The x, y, and z directions of the Cartesian coordinate are along the lattice vectors a, b, and c, respectively. The Fermi energy is set to zero. Calculated (b) Seebeck coefficient $S$ and (c) anomalous Hall conductivity $\sigma_{xy}$ of FeCr$_2$Te$_4$. The calculated $S$ in low temperature shows good agreement with the experiment. The $\sigma_{xy}$ at the Fermi level (zero) is $\sim$ 127 ($\Omega$ cm)$^{-1}$, much larger than the measured value of 22 ($\Omega$ cm)$^{-1}$.}
\label{MTH}
\end{figure}

The derived carrier concentration $n$ is shown in Fig. 3(a). The $n\sim0.5\times10^{21}$ cm$^{-3}$ at 20 K corresponds to $\sim$ 0.04 holes per formula unit, comparable to the value estimated from $q$. Taken into account a weak temperature-dependent $\rho(T)$ [Fig. 1(c)], the estimated $n\sim1.11\times10^{21}$ cm$^{-3}$ from 484 $\mu\Omega$ cm near 100 K points to a mean free path $\lambda \sim$ 0.44 nm. This is comparable to the lattice parameters and is close to the Mott-Ioffe-Regel limit \cite {GunnarsonO}. The AHC $\sigma_{xy}^A$ ($\approx$ $\rho_{xy}^A / \rho_{xx}^2$) is plotted in Fig. 3(b). Theoretically, intrinsic contribution of $\sigma_{xy,in}^A$ is of the order of $e^2/(hd)$, where $e$ is the electronic charge, $h$ is the Plank constant, and $d$ is the lattice parameter \cite{Onoda2006}. Taking $d \approx V^{1/3} \sim 4.3$ {\AA}, $\sigma_{xy,in}^A$ is estimated $\sim$ 900 ($\Omega$ cm)$^{-1}$, much larger than the obtained values in Fig. 3(b). Extrinsic side-jump contribution of $\sigma_{xy,sj}^A$ is usually of the order of $e^2/(hd)(\varepsilon_{SO}/E_F)$, where $\varepsilon_{SO}$ and $E_F$ is spin-orbital interaction energy and Fermi energy, respectively \cite{Nozieres}. The value of $\varepsilon_{SO}/E_F$ is generally less than $10^{-2}$ for metallic ferromagnets. As we can see, the $\sigma_{xy}^A$ is about 22 ($\Omega$ cm)$^{-1}$ at 20 K and exhibits a moderate temperature dependence. This value is much smaller than $\sigma_{xy,in}^A$ $\sim$ 900 ($\Omega$ cm)$^{-1}$, which precludes the possibility of intrinsic KL mechanism. Based on the band structure, as shown in Fig. 4, we obtained the intrinsic AHC as 127 ($\Omega$ cm)$^{-1}$, which is much larger than the measured value too. The extrinsic side-jump mechanism, where the potential field induced by impurities contributes to the anomalous group velocity, follows a scaling behavior of $\rho_{xy}^A = \beta\rho_{xx}^2$, the same with intrinsic KL mechanism. The scaling behavior of $\rho_{xy}^A$ vs $\rho_{xx}$ gives $\alpha \sim 1.1(2)$ by using $\rho_{xy}^A$ = $\beta\rho_{xx}^\alpha$ [Fig. 3(c)], which also precludes the possibility of side-jump and KL mechanism with $\alpha = 2$. It points to that the skew-scattering possibly dominates, which describes asymmetric scattering induced by impurities or defects and contributes to AHE with $\alpha = 1$. Furthermore, the scaling coefficient $S_H = \mu_0R_s/\rho_{xx}^2 = \sigma_{xy}^A/M_s$ [Fig. 3(d)] is weakly temperature-dependent and is comparable with those in traditional itinerant ferromagnets, such as Fe and Ni ($S_H \sim 0.01 - 0.2$ V$^{-1}$) \cite{Dheer,Jan}. It is proposed that the FIM in FeCr$_2$Te$_4$ is itinerant ferromagnetism among antiferromagnetically coupled Cr-Fe-Cr trimers \cite{LIUYu}. In a noncomplanar spin trimer structures the topologically nontrivial Berry phase is induced by spin chirality rather than spin-orbit effect, resulting in chirality-induced intrinsic AHE \cite{OghushiK,ShindouR,TataraG,GaoS}. Our result excludes such scenario in Cr-Fe-Cr trimers in FeCr$_2$Te$_4$ \cite{LIUYu}.

\section{CONCLUSIONS}

In summary, we studied the electronic transport properties and AHE in FeCr$_2$Te$_4$ single crystal. The AHE below $T_c$ = 123 K is dominated by extrinsic skew-scattering mechanism rather than the intrinsic KL or extrinsic side-jump mechanism, which is confirmed by our DFT calculations. The spin structure of Cr-Fe-Cr trimers proposed for FeCr$_2$Te$_4$ is of interest to check by neutron scattering experiments on powder and single crystals in the future.

\section*{Acknowledgements}

Work at Brookhaven National Laboratory (BNL) is supported by the Office of Basic Energy Sciences, Materials Sciences and Engineering Division, U.S. Department of Energy (DOE) under Contract No. DE-SC0012704. B.Y. acknowledges the financial support by the Willner Family Leadership Institute for the Weizmann Institute of Science, the Benoziyo Endowment Fund for the Advancement of Science,  Ruth and Herman Albert Scholars Program for New Scientists, the European Research Council (ERC Consolidator Grant No. 815869, ``NonlinearTopo'').

$^{*}$Present address: Los Alamos National Laboratory, MS K764, Los Alamos NM 87545, USA

\end{document}